\documentclass{elsarta}
\begin{document}
\hspace*{3.5 in}CUQM-111, HEPHY-PUB 808/05\\
\hspace*{3.5 in}math-ph/0508009\\
\hspace*{3.5 in}[August 2005]
\vspace*{0.4 in}
\begin{frontmatter}
\title{Schr\"odinger upper bounds to semirelativistic eigenvalues}
\author[montreal]{Richard L. Hall\corauthref{cor}} and
\corauth[cor]{Corresponding author.}
\ead{rhall@mathstat.concordia.ca}
\author[wien]{Wolfgang Lucha}
\ead{wolfgang.lucha@oeaw.ac.at}
\address[montreal]{Department of Mathematics and Statistics, Concordia University,
1455 de Maisonneuve Boulevard West, Montr\'eal,
Qu\'ebec, Canada H3G 1M8}
\address[wien]{Institut f\"ur Hochenergiephysik, \"Osterreichische 
Akademie der Wissenschaften, Nikolsdorfergasse 18, A-1050 Wien, Austria}
\begin{abstract}
 Problems posed by semirelativistic Hamiltonians of the form $H = \sqrt{m^2+p^2} + V(r)$ are studied. It is shown that energy upper bounds can be constructed in terms of certain related Schr\"odinger operators; these bounds include free parameters which can be chosen optimally. 
\end{abstract}

\end{frontmatter}
\section{Introduction}
We study semirelativistic Hamiltonians $H$ composed of the relativistically
correct expression $K(p^2) = \sqrt{m^2 + p^2},$ $p \equiv |{\bf p}|,$ for
the energy of a free particle of mass $m$ and momentum ${\bf p},$ and of a
coordinate-dependent static interaction potential $V(r),$ $r \equiv |{\bf
r}|,$ which may be chosen arbitrarily, apart from the constraint imposed
on $H$ that it be bounded from below:
$$H = \sqrt{m^2 + p^2} + V(r).\eqno{(1.1)}$$
The eigenvalue equation generated by this kind of Hamiltonian is usually called the spinless Salpeter equation. It arises as a well-defined approximation to
the Bethe-Salpeter formalism for the description of bound states within
(relativistic) quantum field theory \cite{bse} when it is assumed that the bound-state constituents interact instantaneously and propagate like free particles \cite{se}. At the same time, $H$ may be regarded as the simplest and perhaps most straightforward generalization of a (nonrelativistic) Schr\"odinger operator towards the incorporation of relativistic kinematics. For many potentials, this Hamiltonian can be shown \cite{lieb} to be bounded below and essentially self-adjoint, and its spectrum can be defined variationally.  For definiteness, we consider the corresponding eigenvalue problem in three spatial dimensions.

In Sec.~2 we review the well-known tangential Schr\"odinger upper bounds which may be found \cite{halld}-\cite{hallf} either by use of optimized operator inequalities or by the exploitation of the concavity of the Salpeter kinetic-energy $K$ as a function of $p^2.$ The new Schr\"odinger bounds which are the principal concern of this paper are derived by considering operator differences. We shall now illustrate the main idea by considering a nonrelativistic example.  Suppose we wish to estimate the bottom of the spectrum of 
$$H = p^2 +\alpha r^4 -\beta r^2 = H_1 - H_2,\eqno{(1.2)}$$
where $\alpha$ and $\beta$ are positive and
$$H_1 = (1+\omega)p^2 + \alpha r^4,\quad{\rm and}\quad H_2 = \omega p^2 + \beta r^2, \quad \omega > 0.\eqno{(1.3)}$$
Since $H_1 = H + H_2,$ we conclude from the theorem of Weyl \cite{weyl}-\cite{wein} that $E_1 \geq E + E_2.$  We note in passing that for the ground-state energies discussed here, the Weyl inequality follows immediately by applying the exact normalized wave function $\psi_1$ corresponding to $H_1$ as a trial function for the terms of the sum.  Thus we have 
$$E_1 = \langle\psi_1,H_1\psi_1\rangle = \langle\psi_1,H\psi_1\rangle + \langle\psi_1,H_2\psi_1\rangle\geq E + E_2.\eqno{(1.4)}$$
Of course, we assume that the operator domains allow this.  It remains to optimize the expression for $E$ with respect to $\omega > 0.$ Thus we find in the example
$$E \leq \min_{\omega}\left[E_1(\omega)-E_2(\omega)\right] = 
\min_{\omega}\left[e_4\left((1+\omega)^2\alpha\right)^{1/3}-e_2\left(\omega\beta\right)^{1/2}\right],\eqno{(1.5)}$$
where in three dimensions we have for the respective spectral bottoms
$$p^2 + r^4\rightarrow E = e_4 \approx 3.799673,\quad{\rm and}\quad p^2 + r^2 \rightarrow E = e_2 = 3.\eqno{(1.6)}$$
The coupling dependence is found by the general scaling law 
$$p^2 + v\ {\rm sgn}(q)\ r^q\rightarrow E(v) = E(1)\ v^{2/(2+q)}\eqno{(1.7)}$$
 for pure powers $0\neq q > -2.$ For our problem and the special case $\alpha = \beta = 1,$ we find the result
$$2.8345362\approx E < E_u = 2.85525\quad{\rm for}\quad \omega = 0.818584.\eqno{(1.8)}$$
By this reasoning we determine the energy of $H = p^2 +r^4-r^2$ with error less than $ 0.74\%.$ In Sec.~3 we shall show how this idea can be applied to the Salpeter eigenvalue problem.  

\section{Tangential Schr\"odinger upper bounds}
The kinetic-energy term $K(p^2) = \sqrt{m^2+p^2}$ in the Hamiltonian $H = K + V$ is a concave function of $p^2.$ Thus tangents to $K$ generate upper Schr\"odinger operators ${\mathcal H}^{(t)}$ of the form 
$$H \leq {\mathcal H}^{(t)} = a(t)p^2 + b(t) + V(r),\eqno{(2.1)}$$
where $t>0$ is the point of contact between the tangent $ap^2 + b$ and $K(p^2).$  Elementary analysis allows us to obtain the following formulas for the coefficients $a(t)$ and $b(t)$:
$$a(t) = {1\over{2\sqrt{m^2+t}}},\quad b(t) = {{2m^2+t}\over{2\sqrt{m^2+t}}}.\eqno{(2.2)}$$
If an eigenvalue of the Schr\"odinger operator $ap^2 + V(r)$ is given by ${\mathcal E}(a),$ then
we have by the variational characterization of the discrete spectrum of $H$ that the corresponding eigenvalue $E$ of $H$ is bounded by the inequality
$$E \leq \min_{t >0} \left[{\mathcal E}(a(t)) + b(t)\right].\eqno{(2.3)}$$
The minimum in this expression simply picks out the energy of the best upper tangential operator. We have shown earlier \cite{halld} that these `envelope bounds' are identical to those obtained by optimizing over the parameter $\mu$ in the upper bound for $K(t)$ implied by the inequality $\|K(t)-\mu\|^2 \geq 0,$ namely $K \leq (K^2+\mu^2)/(2\mu);$ the link between the two expressions for the bound is the parameter relation $\mu = \sqrt{m^2 + t}.$  

The advantage of the tangential bound is its generality: it applies to each discrete eigenvalue that exists for the upper operator.  For later comparison we consider three examples.  We restrict our attention to the lowest eigenvalue, which is the subject of the difference upper bound discussed in Sec.~3 below. 
\subsection{The ultrarelativistic harmonic oscillator $H = p + r^2~~(m=0)$}~~
By the spectral equivalence $H\equiv\tilde{H} = p^2 + r,$ we see that the exact energy is given by $-z_0$, where $z_0$ is the first zero of Airy's function ${\rm Ai}(z).$  That is to say $E \approx 2.3381074.$ In order to compute the envelope bound we may re-parametrize (2.2) in terms of $s = a = 1/(2\sqrt{t})$ and find $b = 1/(4s).$ The tangential Hamiltonian then becomes ${\mathcal H} = sp^2 + 1/(4s) + r^2.$ The lowest eigenvalue of this operator is then given by ${\mathcal E}(s) = 3\sqrt{s} + 1/(4s).$  If we minimize ${\mathcal E}(s)$ with respect to $s$ we find $E \leq (9/2)6^{-{1\over 3}}\approx 2.47644.$ This is about $5.9\%$ high.
\subsection{The semirelativistic harmonic oscillator $H = \sqrt{1 + p^2} + r^2~~(m=1)$}~~
In this case the Hamiltonian $H$ is equivalent to the Schr\"odinger operator $\tilde{H} = p^2 + \sqrt{1+r^2}$ whose exact energy $E$ is straightforward to find numerically and is given by $E\approx 2.6640196.$  Meanwhile the tangential operator is given by ${\mathcal H} = ap^2 + b + r^2$ and has lowest energy ${\mathcal E}(t) = 3\sqrt{a(t)} + b(t),$ where $a(t)$ and $b(t)$ are given by the formulas (2.2).  A minimization of ${\mathcal E}(t)$ with respect to $t$ yields the best upper bound $E \leq {\mathcal E}(3) = 11/4.$ This bound is about $3.2\%$ high.  As $m$ is increased in the operator $H = \sqrt{m^2 + p^2} + r^2,$ the problem spectrally (and monotonically) approaches the Schr\"odinger limit $H = m + p^2/(2m) + r^2,$ for which the envelope approximation is exact.
\subsection{The ultrarelativistic linear potential $H = p + r~~(m=0)$}~~
This very symmetrical operator is truely non-local but yields easily to a variational treatment in a Hermite basis of the form $\phi(r) = \exp(-\half r^2)\sum_i c_i H_{4i+1}(r).$ In such a basis, each term is form invariant with respect to transformations to momentum space.  The Hamiltonian $H$ has earlier been studied by Boukraa and Basdevant \cite{bouk} with the aid of special methods for solving problems in momentum space. Thus we know that the bottom of the spectrum of $H$ to $4$ places is $E = 2.2322.$  By use of the tangential bound we obtain an upper family of operators of the form ${\mathcal H} = sp^2 + 1/(4s) + r.$ The corresponding lowest eigenvalue is given by ${\mathcal E}(s) = s^{\frac{1}{3}}(-z_0) + 1/(4s),$ where ${\rm Ai}(z_0) = 0,$ and $z_0 \approx -2.3381074.$ By minimizing over $s$ we find the best upper bound to be $E \leq (4/3)\left(3|z_0|^3/4\right)^{\frac{1}{4}}\approx 2.3461,$ that is to say, about $5.1\%$ high.

We shall return to these examples in Sec.~4 and find sharper upper estimates.
\section{Difference Schr\"odinger upper bounds}
The upper bound we shall discuss was discovered in connection with our studies of the semirelativistic many-body problem.  For the $1$-particle case the bound is most easily constructed by means of the following defining equations:
$$
H = \sqrt{m^2 + p^2} + V(r) = H_1 - H_2,\eqno{(3.1)}$$ 
where
\begin{eqnarray*}
H_1 &&= \sqrt{m^2 + p^2} + ap^2 + br^2 \equiv \tilde{H}_1 = bp^2 + \sqrt{m^2 + r^2} + ar^2,\\
H_2 &&= a p^2 + b r^2 - V(r),~~~~~~~~~~~~~~~~~~~~~~~~~~~~~~~~~~~~~~~~~~~~~~~~~~~~~~~~~(3.2)
\end{eqnarray*}
and the parameters $a$ and $b$ are positive. We shall assume that the harmonic oscillator potential $br^2$ dominates the potential $V(r)$ for large $r$.  In this case the operators $\tilde{H}_1$ and $H_2$ are both Schr\"odinger operators whose spectral bottoms we write respectively as $E_1(a,b)$ and $E_2(a,b).$  These energies can of course be found equivalently from eigenproblems expressed in coordinate or momentum space. We let $E$ be the bottom of the spectrum of $H$ and we express the relation between the Hamiltonian operators in the form
$$H_1 = H + H_2.\eqno{(3.3)}$$
It is clear by an elementary variational argument applied to (3.3) that we can conclude the Weyl energy inequality \cite{weyl}-\cite{wein}
$$E_1(a,b) \geq E + E_2(a,b).\eqno{(3.4)}$$
By re-writing (3.4) and optimizing with respect to the free positive parameters $a$ and $b$, we find that our best such difference upper bound to $E$ is given by
$$E \leq E_u = \min_{\{a,\ b\}}\left[E_1(a,b) - E_2(a,b)\right].\eqno{(3.5a)}$$
By adding and subtracting the oscillator $ap^2 + br^2$ in the reverse way we arrive, by exactly similar reasoning, at an alternative difference upper-bound formula given by
$$E \leq E_u^{(-)} = \min_{\{a,\ b\}}\left[E_2^{(-)}(a,b) - E_1^{(-)}(a,b)\right],\eqno{(3.5b)}$$
where the corresponding operators $H_1^{(-)}$ and $H_2^{(-)}$ are defined by
\begin{eqnarray*}
H_1^{(-)} &&= -\sqrt{m^2 + p^2} + ap^2 + br^2 \equiv \tilde{H}_1^{(-)} = bp^2 - \sqrt{m^2 + r^2} + ar^2,\cr
H_2^{(-)} &&= a p^2 + b r^2 + V(r).
~~~~~~~~~~~~~~~~~~~~~~~~~~~~~~~~~~~~~~~~~~~~~~~~~~~~~~~(3.6)
\end{eqnarray*}
Equations (3.5a) and (3.5b) summarize the principal results of this paper.
\section{Examples}
We now consider the three problems mentioned in Sec.~2.  In each case we must solve the corresponding Schr\"odinger problems, $H_1$ and $H_2,$ defined in (3.2), and then minimize the corresponding eigenvalue difference $E_1(a,b)-E_2(a,b)$ with respect to the parameters $a$ and $b$.
\subsection{The ultrarelativistic harmonic oscillator 
$H = p + r^2~~(m=0)$}~~
The corresponding pair of Schr\"odinger operators given by (3.2) become 
$$\tilde{H}_1 = bp^2 + ar^2 + r, \quad H_2 = ap^2 + (b-1)r^2.\eqno{(4.1)}$$
We find from (3.5a)
$$ 2.3381074 \approx E < 2.3433 = 5.63456 - 3.29126 \quad(a = 0.59, b = 3.04).\eqno{(4.2)}$$

\subsection{The semirelativistic harmonic oscillator 
$H = \sqrt{1 + p^2} + r^2~~(m=1)$}~~
The corresponding pair of Schr\"odinger operators are  
$$\tilde{H}_1 = bp^2 + ar^2 + \sqrt{1 + r^2}, \quad H_2 = ap^2 + (b-1)r^2.\eqno{(4.3)}$$
From (3.5a) we find
$$2.6640167\approx E < 2.6689 = 6.33418-3.66528\quad(a = 0.59, b = 3.53).\eqno{(4.4)}$$

\subsection{The ultrarelativistic linear potential $H = p + r~~(m=0)$}~~
The corresponding pair of Schr\"odinger operators are given by 
$$\tilde{H}_1 = bp^2 + ar^2 + r, \quad H_2 = ap^2 + br^2 - r.\eqno{(4.5)}$$
We can derive the best upper bound provided by Eq.~(3.5) analytically in this case.  We find
$$E \leq E_u = \lim_{a\rightarrow\infty}\left[E_1(a,a)-E_2(a,a)\right] = {{(\phi,2r\phi)}\over{(\phi, \phi)}} = {4\over{\sqrt{\pi}}}\approx 2.25676,\eqno{(4.6)}$$
where $\phi(r) = \exp(-\half r^2).$ We can see this by the following argument.  If we let the bottom of the spectrum of the perturbed oscillator $p^2 + r^2 + \lambda r$ be $e(\lambda),$ and we write $a = s^4$ and $b = t^4,$ then by scaling arguments we obtain the equation
$$E_1(a,b) - E_2(a,b) = s^2t^2\left[e\left(1/(s^3t)\right) - e\left(-1/(t^3s)\right)\right].\eqno{(4.7)}$$ 
This expression provides an upper bound for every choice of $s$ and $t.$ The difference will be small when both the expressions for $\lambda$ are small, that is to say, when $s$ and $t$ are large.  In the limit of small $\lambda,$ we know by perturbation theory that the approximation $e(\lambda)\approx 3+(2/\sqrt{\pi})\lambda$ is asymptotically exact.  Thus we find in this small-$\lambda$ limit that 
$$E_1(a,b) - E_2(a,b) \approx {2\over{\sqrt{\pi}}}\left({t\over s} + {s\over t}\right)\geq {4\over{\sqrt{\pi}}}.\eqno{(4.8)}$$
The minimum implies that $s = t,$ and the small-$\lambda$ limit implies that $s\rightarrow \infty.$ Thus the best upper bound provided by the smallest spectral difference is given by the right-hand side of (4.8), as claimed above.

It is evident that the difference upper bound leads to more accurate results for these problems than does the envelope upper bound. We note that the bounds provided by the alternative difference inequality (3.5b) are very similar in numerical quality.

\section{Conclusion}
The main attraction of the Salpeter Hamiltonian $H = \sqrt{m^2 + p^2} + V(r)$  is that it captures some relativistic features whilst remaining a relatively simple operator.  By simple we mean that for many potentials, its spectrum can be defined variationally.  Thus it is in principle straightforward to find energy upper bounds by exploring a finite-dimensional trial space.  The main technical difficulty concerning the Hamiltonian is that, apart from the harmonic oscillator $V(r) = r^2,$ the Hamiltonian is in general non-local.  In the present paper we explore a new class of Schr\"odinger operator differences that provide upper bounds.  The ultrarelativistic linear problem $H = p + r$ shows that in some cases we may expect to obtain analytical results from these bounds.    

 \section*{Acknowledgement}
Partial financial support of this work under Grant No. GP3438 from the 
Natural Sciences and Engineering Research Council of Canada, and the hospitality of the Institute for High Energy Physics of the Austrian Academy of Sciences in Vienna, are gratefully acknowledged by one of us ([RLH]).
\clearpage 

\end{document}